# Golden Bars of Consensus and the Truth Quark

```
    Scientists are imprisoned by Golden Bars of Consensus, says Burton Richter.
              A case in point is the mass of the Truth Quark.
      The consensus analysis of the experimental data indicates that
                the mass of the Truth Quark is about 170 GeV.
    On the other hand, an alternative analysis of the same date indicates that
                the mass of the Truth Quark is about 130 GeV.
  If the design of future experiments takes into account only the consensus value,
              then results of future experiments might be compromised.
```

Burton Richter, in remarks at Stanford in 1999 (hep-ex/0001012) said "... The next big machine, the LHC, is under construction at CERN ... The two main experiments ATLAS and CMS will each have 1500 to 2000 collaborators. ... In the 500-strong collaborations of today, we already have a bureaucratic overlay to the science with committees that decide on the trigger, data analysis procedures, error analysis, speakers, paper publications, etc. The participating scientists are imprisoned by

### golden bars of consensus ....

this will become more difficult as the collaborations grow to three times the size of today's largest. This needs thinking about and talking about, ...". Do the golden bars of consensus already imprison useful data and analysis? Consider, as an example, the

### mass of the T-quark.

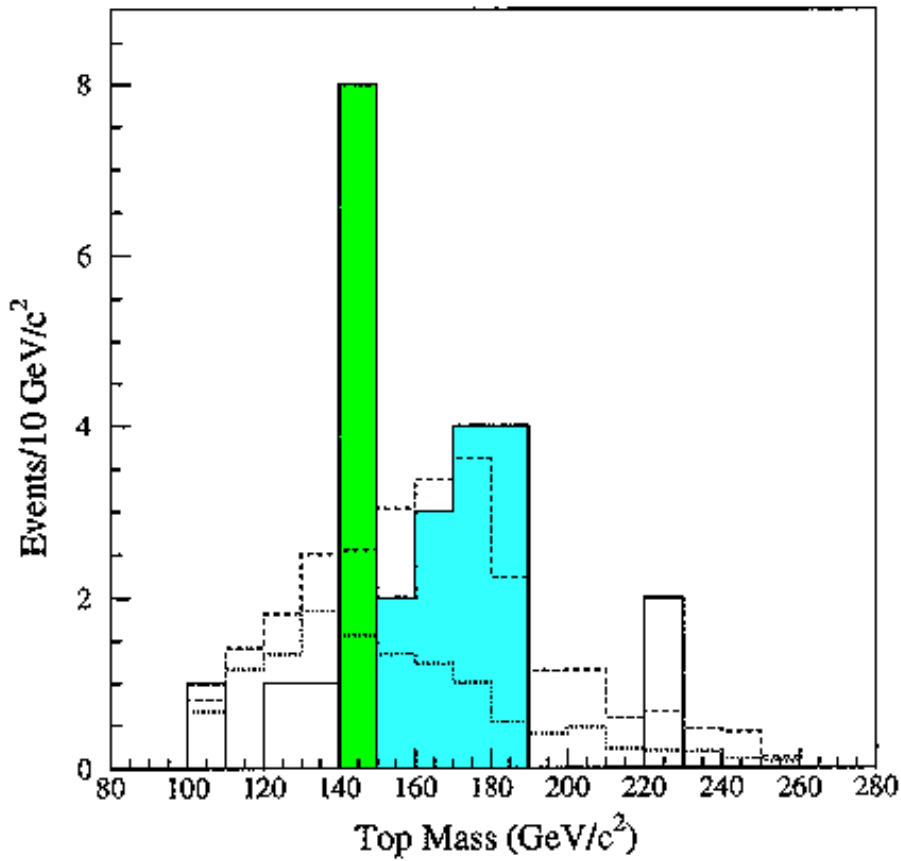

Table of Contents:

- T-quark observations at Fermilab
- semileptonic data - tagged semileptonic data - dileptonic data
- Truth's Consequences
- How is this related to the physics model described at http://www.innerx.net/personal/tsmith/d4d5e6hist.html ?
- What about indirect ElectroWeak determinations?
- What is the 170-180 GeV peak?
- Truth Ceng Zi

In April 1994, CDF at Fermilab (in FERMILAB-PUB-94/097-E) reported a T-quark mass of 174 (+/-10)(+13/-12) GeV. The data analyzed by CDF included a 26-event histogram for W + (3 or more) jets, without b-tags, which is Figure 65 of the report, to which I have added blue and green colors to make discussion easier.

Some of the CDF histogram events, shown in blue, are in the 150-190 GeV range and do support the CDF analysis. However, there is a peak of 8 events in the 140-150 GeV bin, shown in green, that were excluded from the analysis by CDF, saying (on page 140 of the report) "... the bin with masses between 140 and 150 GeV/c^2 has eight events.

<span style="color:red">We assume the mass combinations in the 140 to 150 GeV/c^2 bin represent a statistical fluctuation since their width is narrower than expected for a top signal. ...".</span>

If the 140-150 GeV peak were only a statistical fluctuation seen by the CDF detector, one would not expect to find such a peak repeated in the data seen by the D0 detector at Fermilab. However, in March 1997, D0 (in hep-ex/9703008) reported a T-quark mass of 173.3 GeV (+/- 5.6 stat +/- 6.2 syst), based on data including a histogram similar to Figure 65 of the April 1994 CDF report which is Figure 3 of the D0 report, to which I have added blue and green colors to make discussion easier:

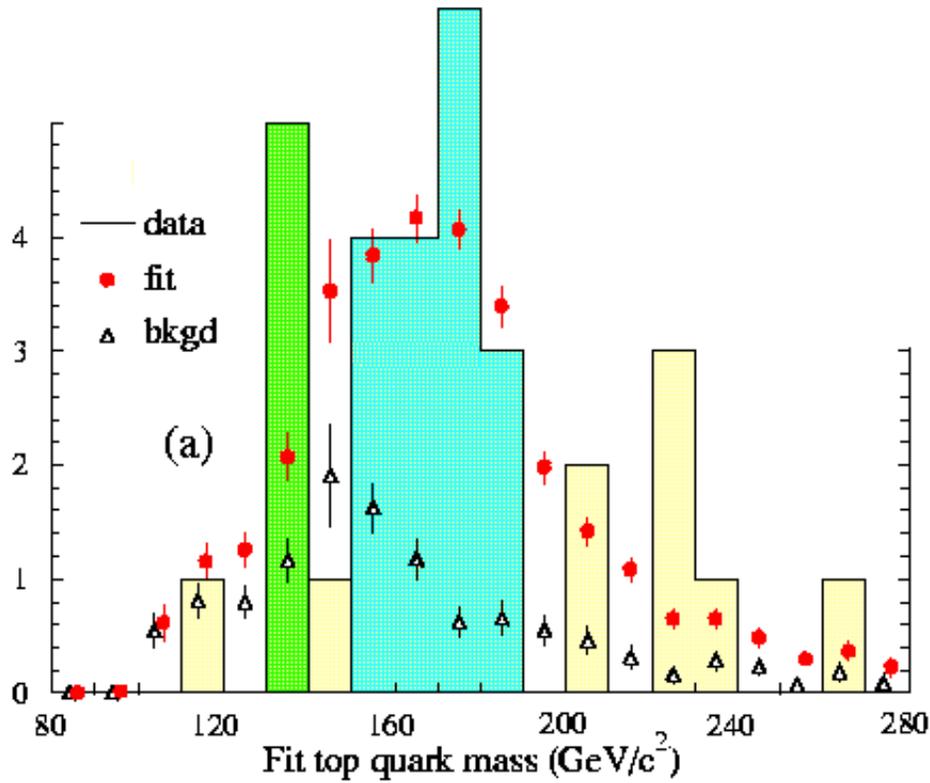

Some of the D0 histogram events, shown in blue, are are in the 150-190 GeV range and do support the CDF analysis. However, similar to the 140-150 GeV bin peak seen and thrown out by CDF, there is a peak of 5 events in the 130-140 GeV bin, shown in green, that were excluded from the analysis by D0. I did not see in the D0 report an explicit discussion of the 5-event peak in the 130-140 GeV bin.

Those 130-150 GeV peaks are from untagged semileptonic events.

---

Tagged semileptonic events may be a more reliable measure of T-quark mass, although there are fewer of them, so that statistics are not as good.

CDF (in hep-ex/9801014, dated 30 September 1997) reported a T-quark mass of 175.9 +/- 4.8(stat.) +/- 4.9(syst.) GeV based on events that were either SVX tagged, SVX double tagged, or untagged. However, CDF analysis of tagged semileptonic events (14 of them) gave a T-quark Mass of 142 GeV (+33, -14), as shown in their Figure 2, which is a plot of events/10 GeV bin vs. Reconstructed Mass in GeV:

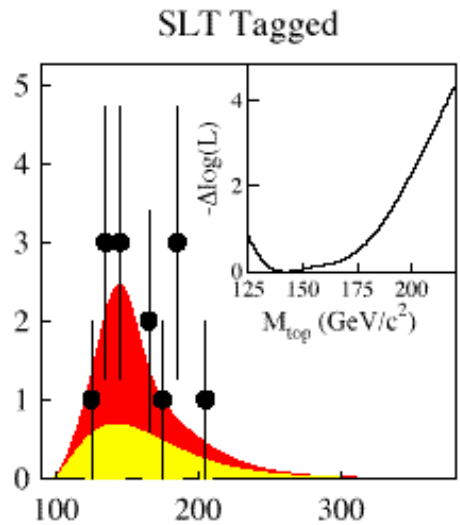

D0 (in hep-ex/9801025) also analyzed tagged semileptonic events, with the result shown in their figure 25:

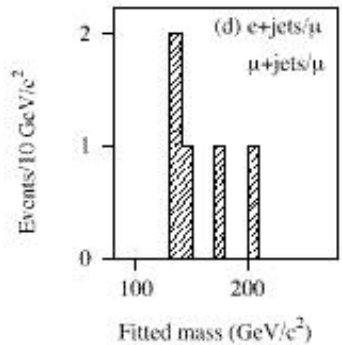

The figure shows 3 events in the 130-150 GeV range, one event in the 170-180 GeV bin, and one event in the 200-210 GeV bin. According to footnote 10 of hep-ex/9801025,

One event which would have otherwise passed the cuts, event (95653; 10822), was removed by D0 from its analysis because it was selected by the dilepton mass analysis. If this event is treated as a l + jets candidate, it has a fit Chi-squared of 0.92 and fitted Truth Quark mass of 138.7 GeV.

---

Dilepton events may be the most reliable measure of T-quark mass, although they are the least numerous type of event, so that statistics are not so good.

In October 1998 (in hep-ex/9810029) CDF analyzed 8 dilepton events and reported a T-quark mass of 167.4

+/- 10.3(stat) +/- 4.8(syst) GeV. Figure 2 of the report shows the 8 events:

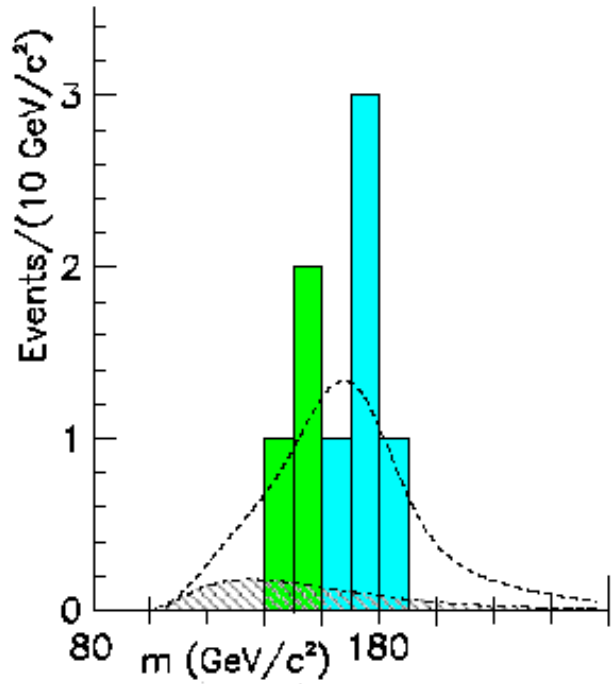

I have colored green the events with T-quark mass less than 160 GeV, and blue the events with T-quark mass greater than 160 GeV. The hep-ex/9810029 CDF report stated that it "... supersedes our previously reported result in the dilepton channel ...".

The superseded previous CDF dilepton report (hep-ex/9802017) analyzed 9 events out of a total of 11 events, which 11 events are shown on the following histogram:

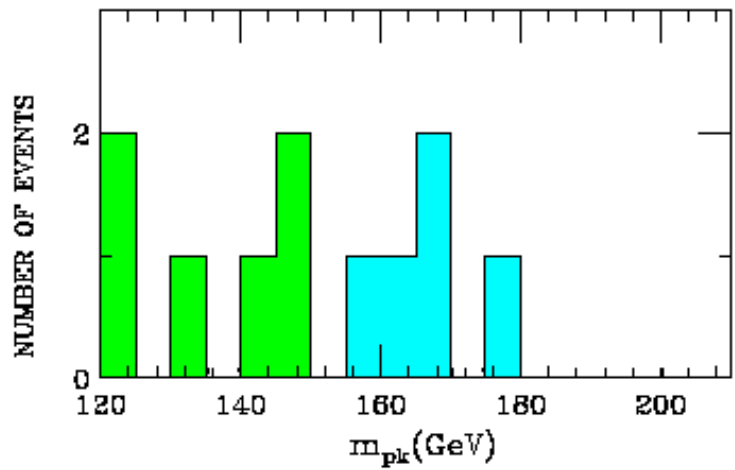

The distribution of $m_{pk}$ values determined from 11 CDF dilepton events available empirically.

I have colored green the events with T-quark mass less than 150 GeV, and blue the events with T-quark mass greater than 150 GeV.

Note first, that in the earlier 11-event histogram 5 events are shown as greater than 150 GeV, but only 4 events are shown as greater than 160 GeV, while in the 8-event revised histogram 5 events are shown as greater than 160 GeV. This indicates to me that some changes in the analysis have shifted the event mass assignments upward by about 10 GeV.

Note second, that the earlier 11-event histogram contains 3 events from 120-140 GeV that are omitted from the 8-event revised histogram.

D0 (in hep-ex/9706014 and hep-ex/9808029) has analyzed 6 dilepton events, reporting a T-quark mass of about 168.4 GeV. The 1997 UC Berkeley PhD thesis of Erich Ward Varnes which can be found on the web at http://wwwd0.fnal.gov/publications_talks/thesis/thesis.html contains details of the events and the D0 analyses. Each of the 6 events has its own characteristics. In this letter I will only discuss one of them, Run 84676 Event 12814, an electron-muon dilepton event. This figure from page 159 of the Varnes thesis, shows a T-quark mass likelihood plot calculated by the neutrino weighting algorithm.

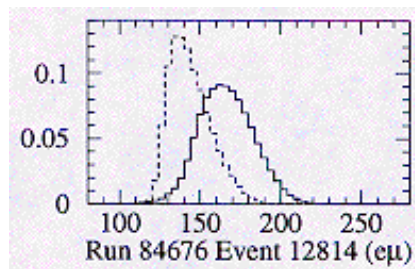

In this event there were 3 jets instead of the 2 jets you would normally expect in a Dilepton event.

The solid line is the plot if all 3 jets are included, and the dashed line is the plot if only 2 of the jets are included by excluding the third (lowest transverse energy) jet.

The 3-jet interpretation supports the 170 GeV mass favored by the Fermilab consensus, while the 2-jet interpretation supports a 130-140 GeV mass analysis that may be imprisoned by the golden bars of Fermilab consensus.

# Could the golden bars of consensus do any real harm to physics ?

## Consider the work going on now in planning experiments to be done on future machines, such as the LHC.

A recent paper (hep-ph/0002205), The Standard Model Instability and the Scale of New Physics, by J.A. Casas, V. Di Clemente and M. Quiros, concludes that "... with the present lower bounds on the Higgs mass, the new physics could easily (but not necessarily) escape detection in the present and future accelerators. ...".

They state "...In this paper we will fix Mt to its experimental mean value and disregard the effect due to the experimental error deltaMt. ...".

If Fermilab had not bound Casas, Di Clemente, and Quiros within the golden bars of consensus of T-quark mass Mt at about 175 GeV, then their paper might have included effects of different values of Mt.

In other words,

restriction of consideration of T-quark masses to around 175 GeV could produce a much more pessimistic view of the likelihood of success of new physics searches at LHC, while breaking the golden bars of consensus to give freedom to consideration of T-quark masses well below 175 GeV produces a much more optimistic view of the likelihood of success of new physics searches at LHC.

Although the economies contributing to the LHC are now prosperous enough to say that it will be built, things can change (and, as was the case for the SSC, not always for the better). Therefore,

I don't think that it is a good idea to imprison ideas that can lead to optimism about the potential for discovery at the LHC.

Since the T-quark is a key part of the truth of nature that physics seeks to understand,

I prefer to call it the Truth quark

(although I realize that now I am in a tiny minority, as most now call it Top).

---

## How are the Golden Bonds of Consensus related to my D4-D5-E6-E7-E8 VoDou Physics model ?

If the Truth quark mass is about 130 GeV, it is consistent with the prediction of my D4-D5-E6-E7-E8 VoDou Physics model made in 1984, shown as the dark blue line:

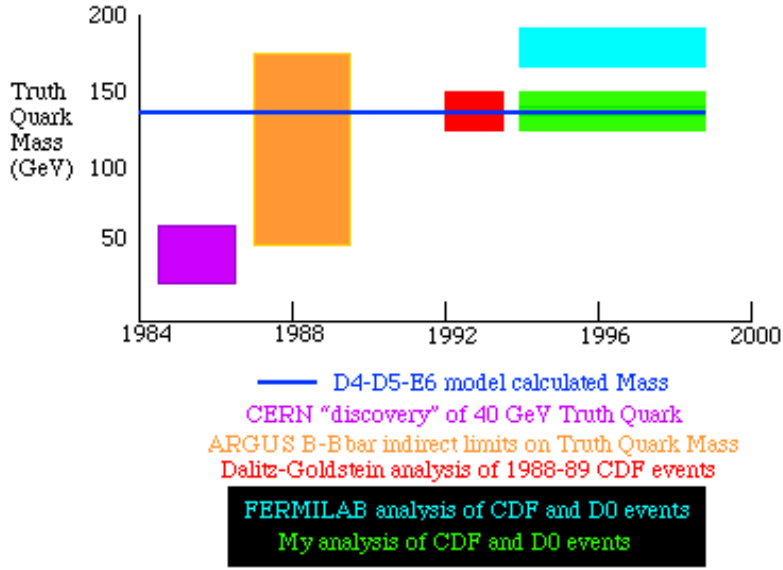

My D4-D5-E6-E7-E8 VoDou Physics model prediction was made before CERN announced in 1984 the "discovery" of the Truth quark with mass 40 GeV, shown as purple, and that during the time that the CERN announcement was generally considered to be valid, I still maintained that CERN was wrong and that my prediction was right.

My D4-D5-E6-E7-E8 VoDou Physics model prediction was consistent with the ARGUS indirect B-Bbar experimental determination, shown as gold; and with the Dalitz-Goldstein analysis of 1988-89 CDF events, shown in red.

The green bar represents my analysis of Fermilab's Run 1, and the cyan bar represents Fermilab's analysis of that data.  **What about indirect electroweak experimental determinations of the Truth quark mass?**

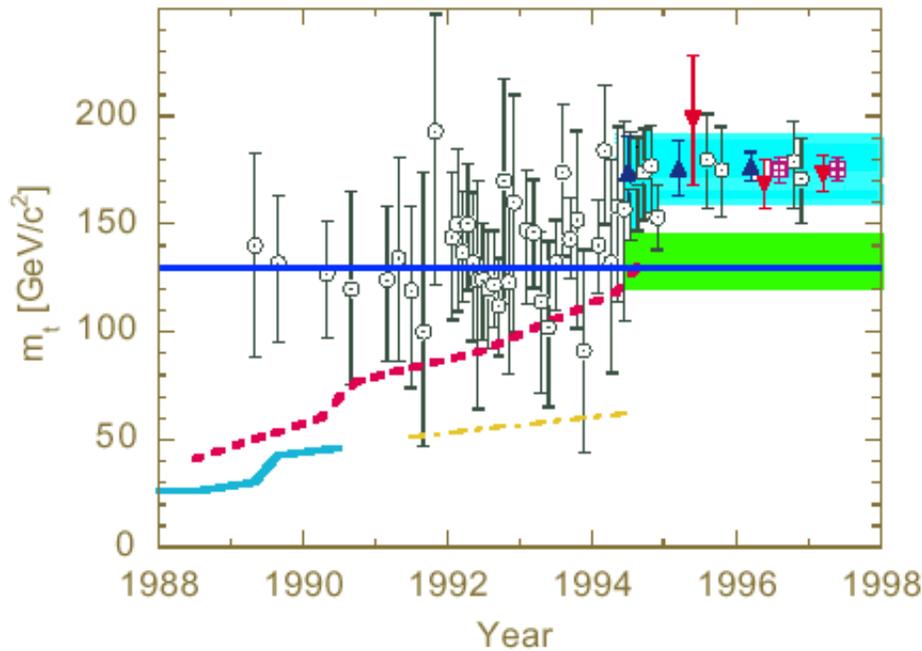

The chart above, from Figure 6 of Chris Quigg's paper hep-ph/0001145, shows "... **Indirect determinations of the top-quark mass from fits to electroweak observables (open circles)** and 95% confidence-level lower bounds on the top-quark mass inferred from direct searches in e + e annihilations (solid line) and in pbar-p collisions, assuming that standard decay modes dominate (broken line). An indirect lower bound, derived from the W-boson width inferred from pbar-p to (W or Z) + anything, is shown as the dot-dashed line. Direct measurements of mt by the CDF (triangles) and D0 (inverted triangles) Collaborations are shown at the time of initial evidence, discovery claim, and at the conclusion of Run 1. The world average from direct observations is shown as the crossed box. ...". I have added the blue line at 130 GeV, and the color of cyan for the region or Fermilab analysis of CDF and D0 events and the color of green for the region of my analysis of them.

## If the 170-180 GeV peak is not the Truth quark, then what is it ?

I don't know for sure, but maybe it is poorly understood background, perhaps related to miscounting the number of jets associated with events, and/or perhaps related to some phenomena seen at HERA and CERN.

According to hep-ex/9910012 by the H1 and ZEUS Collaborations at HERA: "...Between mid-1994 and the end of 1997 the electron-proton collider HERA at DESY has been operated ... . In this period, ZEUS and H1 have collected e + p data samples corresponding to integrated luminosities of 47.7 pb^(-1) and 37 pb^(-1), respectively. In 1998 and the first half of 1999, each experiment has taken about 15 pb^(-1) of ... data ... The excess in the H1 data is still present at Me = 200 GeV but has not been corroborated by the 1997 data. Also ZEUS observes an excess at Mej > 200 GeV; however, the decay angular distribution does not support a LQ interpretation ...

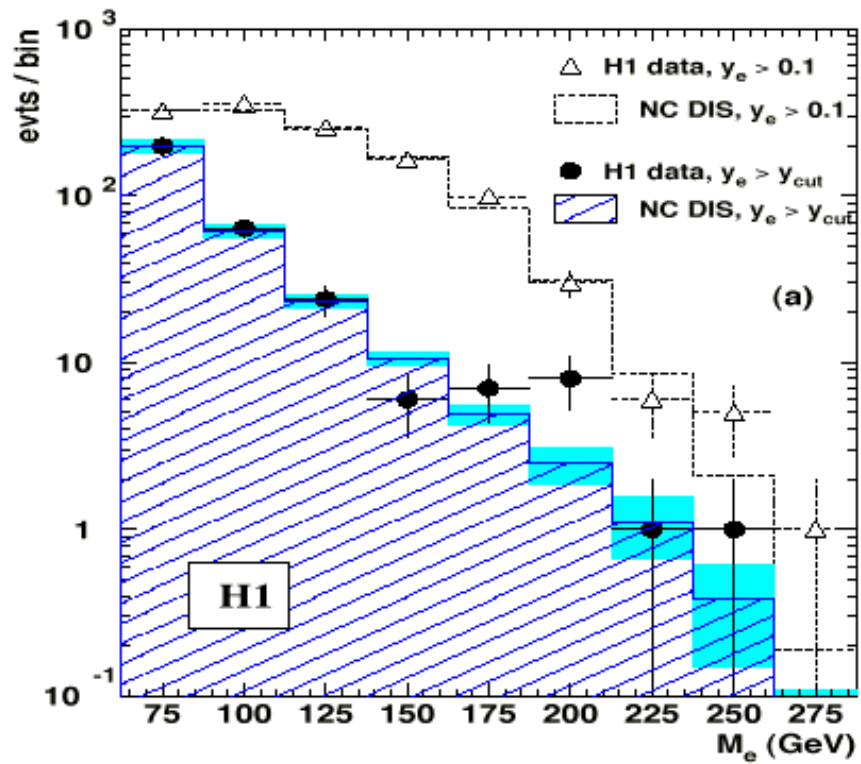

... The symbols with statistical error bars represent the data, the histograms the DIS MC. The H1 data are shown before (open triangles) and after (full dots, hatched histogram) a cut y>ymin(Me). The shaded area indicates the uncertainty of the SM prediction. ...".

In hep-ex/0001014, **Recent Results from the L3 Experiment,** Salvatore Mele of CERN says, [with some comments by me set off by brackets]: "... A data sample corresponding to an integrated luminosity of 232 pb^(-1) was collected in 1997 and 1998 by the L3 experiment at LEP in e+ e- collisions at centre-of-mass energies between 181.7 GeV and 188.7 GeV. ... the cross sections for single W production ... [are compared in this figure] ...

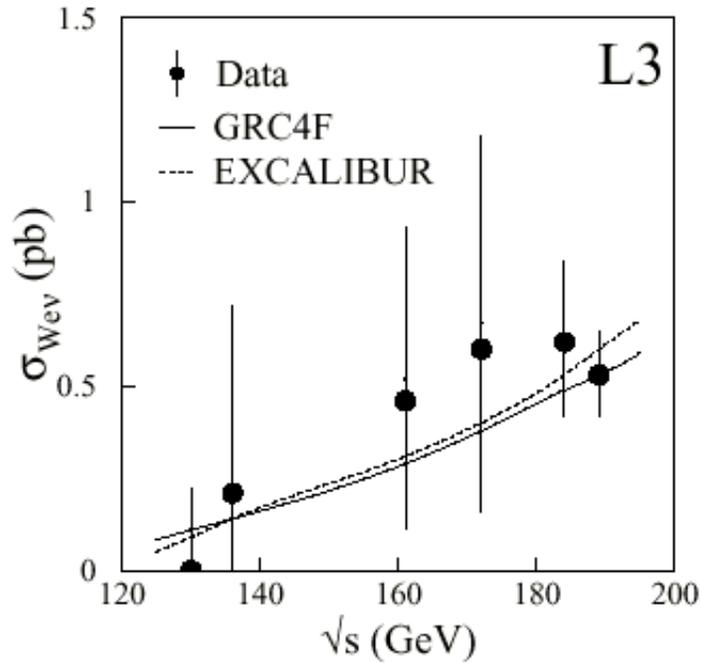

... to the SM predictions obtained with the EXCALIBUR and GRC4F MC [Monte Carlo] programs. ...".

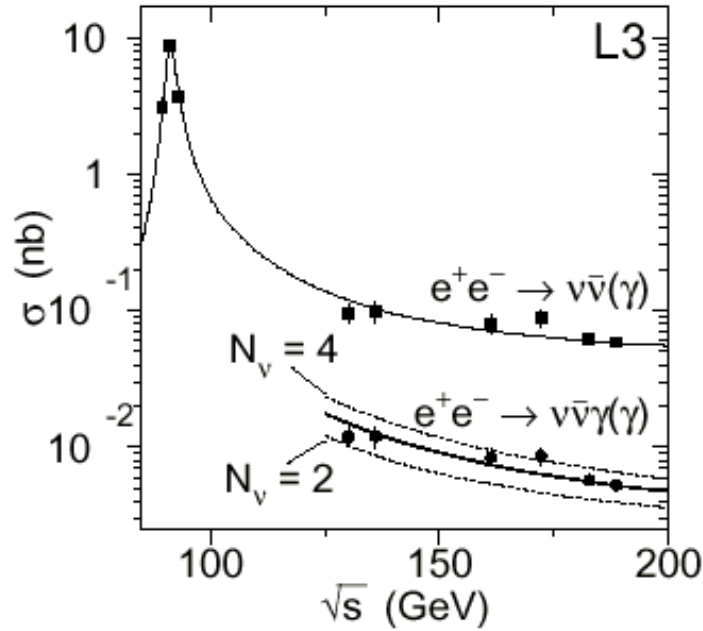

The data in the 175 GeV region is consistent with 4 generations of neutrinos, rather than 3, also indicating to me that processes in that region are incompletely understood.

[note added 25 Feb 2002: Late last night (Sunday 24 Feb 2002) I first viewed an e-mail message sent to me by Salvatore Mele, a copy of which message is quoted here in full:

```
" Delivered-To: tsmith@innerx.net
X-Authentication-Warning: pcl3ep02.cern.ch: smele owned process doing -bs
Date: Sat, 23 Feb 2002 16:43:41 +0100 (CET)
From: Salvatore Mele <Salvatore.Mele@cern.ch>
X-Sender: smele@pcl3ep02.cern.ch
To: tsmith@innerx.net
Subject: Your interpretation of hep-ex/0001014
MIME-Version: 1.0

Dear Mr. Smith,

you authored both the following paper and the following web page:

http://arXiv.org/abs/physics/0006041
http://www.innerx.net/personal/tsmith/TCZchron.html#Jan00-1

I appreciate your reference to the results I summarise in
hep-ex/0001014. I have to draw your attention to the fact that in
experimental physics we do quote an uncertainty associated to measurments.
In the three plots you've extracted from my publication, this is
represented by the vertical bars, called error bars. These concepts are
described in introductory textbooks on statistics.

 Physics results have to be interpreted within these
uncertainties, conventionally associated to a probability of 68% that the
true value of the measured quantity lies in that range.

In this spirit, the first and last of my plots are in excellent agreement
with the predictions and your statement "substantially higher" is void of
any statistical meaning and simply wrong.

As for the neutrino plot, in experimental physics you claim a deviation
from a model when your data are three lenghts of those error bars away
from your prediction that means more than 99% probability of an
inconsistency. This is cleary not the case. In addition, to claim
discoveries or lose faith into a model, five of those error bar lengths
are needed, corresponding to probabilities of the order of 10^-5

Regards,

Salvatore Mele
Research Staff
CERN - European Organisation for Nuclear Research "
```

My comments on Salvatore Mele's message quoted above are:

- To the accuracy of the 1-standard-deviation error bars shown, it is true the error bars are indeed so big that Salvatore Mele's interpretation is indeed not refuted by the plotted experimental results. However, it is also true that my interpretation is not so refuted either. Therefore, I strongly advocate more experimental work so that a distinction can be made.
- My interpretation is consistent with the raw data points. In my opinion, a properly done statistical evaluation would in fact favor my interpretation, although the degree to which that data favors my interpretation would be very slight, as the data deviates from the model preferred by Salvatore Mele by

- only (rough guess on my part) no more than about 1/3 or 1/2 of a standard deviation. I do not claim that such a small deviation is anywhere near being conclusive, but I do claim that, small though it may be, it is not "void of any statistical meaning and simply wrong". However, I do agree with Salvatore Mele that my use of the words "substantially higher" was incorrect, and I should have said "somewhat higher".
- Since the experimental data of Salvatore Mele is only one element that I have considered in evaluating my interpretation, and also probably only one element that Salvatore Mele has considered in evaluating his interpretation, I advocate considering the other elements (my other elements are set out in the balance of this paper/web-page and my other papers/web-pages directly or indirectly referenced or linked thereto) in doing any evaluation of either interpretation. In other words, I like the spirit of Bayesian statistics. I work alone with no staff or outside funding, and have not had the time to do such an analysis, but I would be very interested in seeing a properly done Bayesian evaluation of my interpretation vs. Salvatore Mele's interpretation. Perhaps some day such a study might be done by some large well-funded entity, such as for example CERN. ]

---

Frank D. (Tony) Smith, Jr.

2000 - 5760 - Year of Metal Dragon [note added 2002]

http://tony.ai

http://www.innerx.net/personal/tsmith/TShome.html

The first paper I ever put on the Los Alamos e-print archive, hep-ph/9301210 - Calculation of 130 GeV Mass for T-Quark, was at MENTOR.LANL.GOV for years:

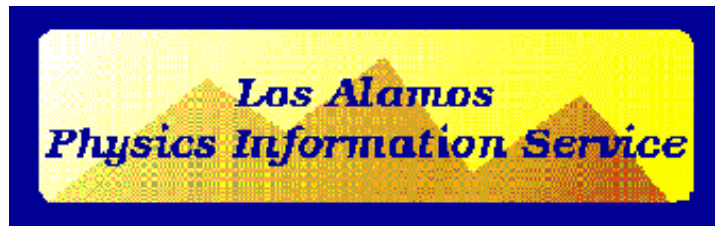

A few people. including me, who like the concepts of Truth and Beauty, still call the T-Quark the Truth Quark instead of the Top Quark.

I also call it Truth Ceng Zi, because Ceng Zi means quark in the language of the physicist who first proposed it, Liu Yao-Yang.

Around 1960, he invented the Ceng Zi (quark) model of particle physics. He was working at the University of Science and Technology of China, which was then located at Beijing, when he invented the Ceng Zi model. He wrote a paper and submitted it to a Chinese journal. It was turned down because the editors thought the paper was not correct. After the quark model had been independently re-invented around 1962-1964, with most of the credit going to Murray Gell-Mann, the editors apologized for rejecting the paper. Liu Yao-Yang is, the last that I heard, still working at the University of Science and Technology of China, which is now at Anhui, in the fields of atomic and molecular physics, quantum field theory, and quantization of gravity.

---